\documentclass{article}
\usepackage{spconf,amsmath,graphicx,amssymb}
\usepackage{stfloats,array,pifont,multirow,multicol,booktabs}
\usepackage{enumitem}
\setlist{nosep, leftmargin=14pt}

\usepackage{mwe} 

\title{Mining fMRI Dynamics with Parcellation Prior\\ for Brain Disease Diagnosis}
\name{\normalsize Xiaozhao Liu$^{1,2,3}$, Mianxin Liu$^{3,4}$, Lang Mei$^{1,2,3}$, Yuyao Zhang$^1$, Feng Shi$^2$, Han Zhang$^3$, Dinggang Shen$^{2,3,5,\ast}$\thanks{*Corresponding author.}}
\address{ \normalsize $^1$School of Computer Science and Technology, Shanghaitech University, China\\
\normalsize $^2$Department of Research and Development, Shanghai United Imaging Intelligence Co., Ltd., China\\
\normalsize $^3$School of Biomedical Engineering, Shanghaitech University, China\\
\normalsize $^4$Shanghai Artificial Intelligence Laboratory, China\\
\normalsize $^5$Shanghai Clinical Research and Trial Center, China
}
\begin{document}

\maketitle

\begin{abstract}

To characterize atypical brain dynamics under diseases, prevalent studies investigate functional magnetic resonance imaging (fMRI). However, most of the existing analyses compress rich spatial-temporal information as the brain functional networks (BFNs) and directly investigate the whole-brain network without neurological priors about functional subnetworks. We thus propose a novel graph learning framework to mine fMRI signals with topological priors from brain parcellation for disease diagnosis. Specifically, we 1) detect diagnosis-related temporal features using a ``Transformer'' for a higher-level BFN construction, and process it with a following graph convolutional network, and 2) apply an attention-based multiple instance learning strategy to emphasize the disease-affected subnetworks to further enhance the diagnosis performance and interpretability. Experiments demonstrate higher effectiveness of our method than compared methods in the diagnosis of early mild cognitive impairment. More importantly, our method is capable of localizing crucial brain subnetworks during the diagnosis, providing insights into the pathogenic source of mild cognitive impairment. 

\end{abstract}
\begin{keywords}
Graph neural network, Brain disease, mild cognitive impairment, Transformer, Multiple instance learning
\end{keywords}
\section{Introduction}
\label{sec:intro}

Functional connectivity (FC) from functional magnetic resonance imaging (fMRI) is the most prevalent measurement to quantify the abnormal brain regional interactions in brain disease studies. Traditional studies apply machine-learning algorithms as feature extractors and classifiers on FC or FC-derived brain functional networks (BFNs) to perform diagnoses \cite{van2010exploring}. The latest works implement graph neural networks (GNNs) to more properly process topological information and to provide more accurate results \cite{liu2021building}\cite{li2021braingnn}\cite{xing2021ds}\cite{gadgil2020spatio}. However, we argue that two limitations in previous works have not drawn enough attention.

First, the diagnosis-related temporal fluctuations from the fMRI regional time series are not optimally utilized. Existing efforts either calculate some indicators manually or use the non-targeted encoder to deal with regional signals as the sequential data \cite{mahmood2021deep}. As a state-of-the-art network architecture, Transformer \cite{vaswani2017attention} has recently demonstrated great capability of extracting temporal features, which can be used as a part of the more optimized encoder to capture brain dynamics.

Second, current studies often input the whole BFN for feature selection, while recent neurological studies suggest brain diseases affect specific brain subnetworks (subnets) \cite{bozoki2012disruption}. Neglecting such important priors, the conventional methods directly search the whole BFN with superfluous parameters, which may lead to over-fitting risks and limited diagnostic performance. The prior also implies that brain disease diagnosis can be formulated as a “multiple instance learning (MIL)” problem, where evidence shown in more than one positive instance (part of the BFN) can support a positive decision. Among all MIL methods, the attention-based MIL \cite{ilse2018attention} is one of the representatives, providing precise decisions and interpretable features.  

To address the above-mentioned issues, we propose a graph learning framework, called “SLMIL-GCTrans”. There are two innovations: 1) We specially design a module that combines GCN and Transformer (named “GCTrans”) to most appropriately extract spatial-temporal features; 2) We utilize the prior knowledge provided by a well-established brain parcellation and apply the attention-based MIL to highlight disease-affected subnets (a subnet-level MIL learning strategy, named “SLMIL”). In this study, we take early mild cognitive impairment (EMCI), known as an early stage of Alzheimer's disease, as a representative brain disease and utilize the data from the Alzheimer’s Disease Neuroimaging Initiative (ADNI) \cite{aisen2015alzheimer} dataset to test our proposed method.


\section{Methods}
\label{sec:methods}

\begin{figure*}[t]
\centering
\includegraphics[width=0.8\textwidth]{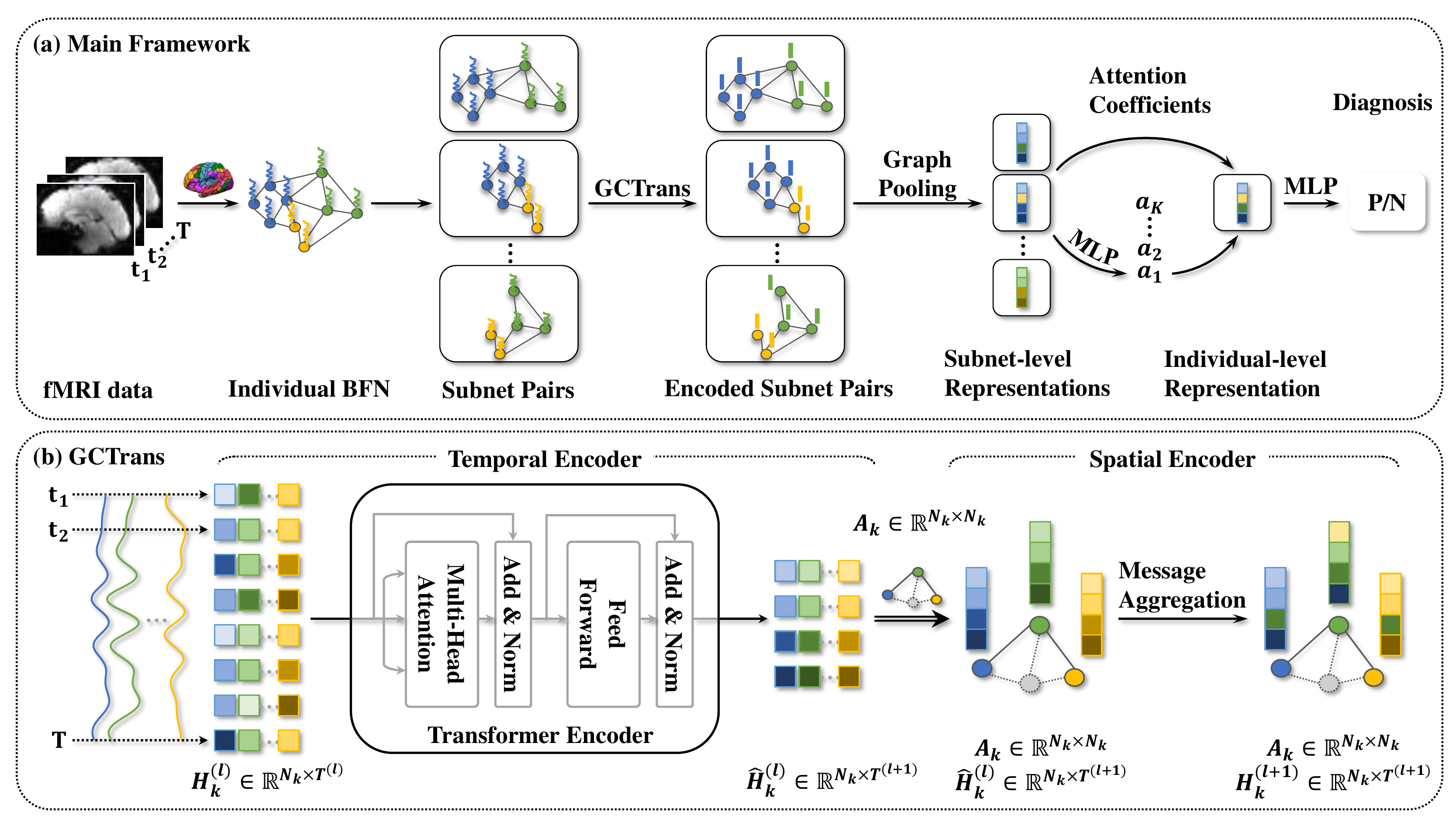}
\caption{An illustration of our proposed SLMIL-GCTrans. (a) The main framework. (b) The detailed design of GCTrans. Subnet instances (i.e., subnetwork pairs) are first generated from the entire BFN. The instances are inputted to GCTrans modules to extract spatial-temporal features. After graph pooling, an attention-based MIL is performed to highlight crucial subnetworks and fuse their features, based on which the categorical diagnosis result is generated.} 
\label{fig1}
\end{figure*}

The entire pipeline of proposed framework is shown in Fig.\ref{fig1}. 


\subsection{Construction of Individual BFN}

From individual fMRI data, we apply a functional parcellation 
 from Schaefer et al. \cite{schaefer2018local} to define brain “regions of interest” (ROIs). The ROIs in Schaefer's parcellation could correspond to 7 widely-recognized functional subnetworks, which can be used as a prior knowledge. Then, the regional averaged time series within each ROI are obtained, which is denoted as $H\in \mathbb{R}^{N\times T}$ ($N$ is the total number of ROIs, and $T$ is the total number of time points). Next, the FC is computed as Pearson correlation between each pair of the ROIs. When computing the FCs among all possible pairs of the ROIs, a BFN can be yielded with the ROIs as nodes and the FCs as edges. Mathematically, BFN can be represented as the adjacency matrix $A\in \mathbb{R}^{N\times N}$. Both BFNs and ROI signals (denoted as $(A,H)$) are used as input for the deep-learning processing.

\subsection{GCTrans Module}
We propose the spatial-temporal encoder “GCTrans”, which aims to integrate the GCN and Transformer to better extract features from both BFNs and ROI signals.

The original GCN is the most prevalent and representative GNN. It computes the graph embedding within each layer with a two-part structure: 1) ``Message Generation'' by processing node features with a multi-layer perceptron (MLP), and 2) ``Message Aggregation'' to nodes from their graph neighbors by multiplying node features with the graph Laplacian. However, the GCN alone is not capable of processing the spatial-temporal BFNs during message generation for its built-in MLP module is not optimized for processing sequential data. 
Among all the sequential models, Transformer is considered a powerful option. Its unique self-attention mechanism allows a learning on feature selection at all time points and the complex inter-time-point correlations. It thus can effectively capture temporal dependencies at any temporal scales and optimally represent and integrate the temporal features.
Therefore, we design GCTrans module (Fig.1b) based on the Transformer encoder to deal with the regional time series to reveal ``brain states'', i.e., a typical hidden state under collective patterns of all ROIs at a given time point.  

Technically, our GCTrans module follows the two-part design of GCN. First part is the ``temporal encoder'' for message generation, where we replace the MLP module with a Transformer encoder. In this part, we compute the similarity and the importance of the collective patterns at all time points $H^{(l)}$ for encoding them into a shorter sequence to describe a more general and higher-level fluctuation $\widehat{H}^{(l)}$. $H^{(l)}$ is the node feature at l-th layer. 
The second part is the ``spatial encoder'' for message aggregation, where we multiply the learned $H^{(l)}$ with the adjacency matrix $A$. Finally, by going through an activation function $Tanh$, we can obtain a graph embedding $H^{(l+1)}$, with the spatial-temporal features properly encoded. The update function of our proposed GCTrans module at layer $l$ can be formulated as follows:  
\begin{align}
    H^{(l+1)}=Tanh(A\mathcal{T} (H^{(l)})),
\label{eq1}
\end{align}
 where $\mathcal{T}(\cdot)$ is the Transformer encoder. In this study, we implemente the classical design of Transformer according to \cite{vaswani2017attention}, which contains a multi-head attention mechanism, a feed-forward layer and the layer normalizations.

\begin{table*}[!ht]
\centering
\caption{Comparative Experiments}
\begin{tabular}{c m{1.2cm}<{\centering} m{1.2cm}<{\centering} m{1cm}<{\centering} m{1cm}<{\centering} c c c} 
\hline
\multirow{2}{*}{Method} & \multicolumn{2}{c}{Modules}       & \multicolumn{2}{c}{Input} & \multirow{2}{*}{Params} & \multirow{2}{*}{AUC} & \multirow{2}{*}{ACC}  \\ 
\cline{2-5}
                         & Spatial encoder & Temporal encoder & FC & ROI signals           &                         &                      &                       \\ 
\hline
GCN                    & GCN  & - & \checkmark & - & 5.9M  & 0.671$\pm$ 0.049 & 0.677$\pm$ 0.048  \\ 
GAT                    & GAT  & - & \checkmark & - & 6.0M  & 0.663$\pm$ 0.045 & 0.651$\pm$ 0.051  \\  
LSTM                     & -  & LSTM & - & \checkmark & 164.0M  & 0.699$\pm$ 0.042 & 0.701$\pm$ 0.039  \\   
TE                       & -  & TE & - & \checkmark & 107.6M  & 0.709$\pm$ 0.039 & 0.712$\pm$ 0.035  \\ 
\hline
GC-LSTM                  & GCN  & LSTM & \checkmark & - & -  & 0.703$\pm$ 0.045 & 0.705$\pm$ 0.040  \\ 
ST-GCN                 & GCN  & - & \checkmark & \checkmark & -  & 0.715$\pm$ 0.039 & 0.717$\pm$ 0.032  \\ 
\textbf{GCTrans(ours)}  & GCN  & TE & \checkmark & \checkmark & 246.0M  & 0.720$\pm$ 0.032 & 0.726$\pm$ 0.034  \\
\textbf{SLMIL-GCTrans(ours)}  & GCN  & TE & \checkmark & \checkmark & 84.3M  & \textbf{0.729$\pm$ 0.022} & \textbf{0.731$\pm$ 0.030}  \\
\hline
\label{tab1}
\end{tabular}
\end{table*}

\subsection{SLMIL Framework}

Our implemented brain parcellation provides information about the 7 functional subnetworks. We then utilize such prior knowledge to generate subnet-level instances. First, we select out the subnet from the entire BFN. Further, by combing different subnets, we can define $K$ instances $\{(A_k,H_k) | k=1,2,...,K\}$ at different spatial scales. Here, $K=C_7^i$ is the number of instances and $i$ is the number of subnets included in each instance. We thus have 6 types of instance generation methods for $i=1,2,...,6$ to characterize functional interactions in different spatial scales. For example, when $i=1$, we directly use each of the 7 separated subnets as 7 instances. When $i=2$, we combine every 2 subnets into an instance, each containing both intra- and inter-subnet interactions of that subnet pair. This method can force the graph encoder to focus on the subnet-level features rather than the entire BFN, therefore reducing the input size and the parameter number of our model. As a hyper-parameter of our method, we have discussed the 6 types of the instance generation methods in Sec.\ref{sec:defferent instances}.

After generating multiple subnet-level instances, processing them by two layers of GCTrans and the graph pooling, we can obtain a set of the embedded vectors $\{H_k^{sub} | k=1,2,...,K\}$ as the subnet-level representations. Subsequently, the attention-based MIL \cite{ilse2018attention} is used to dynamically assign importance $a_k$ to the input instances based on their features. The weights of importance are used for pooling the features of instances, called “attention coefficients”, based on which the diagnosis is generated. In this study, we train an MLP to learn attention weights. The individual-level representation can thus be yield:  
\begin{align}
    H^{ind}=\sum_{k=1}^K a_k H^{sub}_k
\label{eq2}
\end{align}  
where:  
\begin{align}
    a_k=\frac{exp\{\mathcal{MLP}(H_k^{sub})\}}{\sum_{j=1}^K exp\{ 
 \mathcal{MLP}(H_j^{sub})\}}
\label{eq3}
\end{align}

\section{Experiments and Results}
\label{sec:exp}

We first perform the comparative experiments in Sec.\ref{exp:GCTrans} to demonstrate the effectiveness of our proposed modules. Sec.\ref{exp:SLMIL} further discusses the efficiency improvement and the interpretability brought by the SLMIL framework.

\subsection{Data Description}

We employ longitudinal fMRI data from ADNI, which also includes structural MR images as widely used in AD studies \cite{liu2015inherent}\cite{zhang2017alzheimer}. The selected dataset contains 345 EMCI scans and 468 normal control scans. Given that each subject has multiple scans, we adopt a cross-validation scheme in which subjects (instead of the scans) are randomly split 10 times for training and validation.

\subsection{Comparative Experiments}
\label{exp:GCTrans}

We first compare our proposed \textbf{GCTrans} module with other spatial encoders. Among all the prevalent GNNs, we choose and investigate the most representative and widely used 2-layer models of \textbf{GCN} and graph attention network (\textbf{GAT}).The GAT uses the FCs as node features and learns the edges from them.  Specially, both of the models take the entire BFN as input (without generating instances). As prevalent works did, we not only use FCs as the edges but also use an one-to-all FC vector as the node feature for each node in GCN. In this way, the information in ROI signals are neglected.
For the temporal encoder, we implement \textbf{LSTM} and Transformer encoder (\textbf{TE}) to mine ROI signals rather than BFNs and to set another set of baselines.   
Finally, we compare our entire model \textbf{SLMIL-GCTrans} with two related methods. The \textbf{GC-LSTM} from \cite{xing2021ds} and the \textbf{ST-GCN} from \cite{gadgil2020spatio} both attempted to construct the dynamical BFN to incorporate the spatiotemporal information. The GC-LSTM learns on the fluctuations of dynamical BFNs; The ST-GCN use the ROI signals as node feature but do not use any specific design on the temporal encoder.  

As shown in Table \ref{tab1}, our proposed \textbf{GCTrans} module (also with 2 layers) offers higher performance than all of the compared methods. In addition, with SLMIL framework integrated, our complete model \textbf{SLMIL-GCTrans} reduces the number of parameters by 40\% while achieving the best performance. The detailed settings of SLMIL are described in the following section.

\subsection{Performance Improvement and Feature Interpretability under the SLMIL Framework}
\label{exp:SLMIL}

\subsubsection{Performance Improvement}
\label{sec:defferent instances}

\begin{table*}[!ht]
    \centering
    \small
    \caption{Different Ways of Generating Instances}
    \begin{tabular}{m{1.5cm}<{\centering} m{1.8cm}<{\centering} m{1.8cm}<{\centering} m{1.5cm}<{\centering} c c c}
        \hline
        Num of subnet & Intra-subnet information & Inter-subnet information   & Num of instances & Params & AUC & Accuracy    \\
        \hline
        1 & \checkmark & - & 7 & 49.8M &  0.643$\pm$0.055   & 0.651$\pm$0.062 \\
        \textbf{2}  & \textbf{\checkmark} & \textbf{\checkmark} & 21 & \textbf{84.3M} & \textbf{0.729$\pm$0.022} & \textbf{0.731$\pm$0.030}\\
        3 & \checkmark & \checkmark & 35 & 114.7M &  0.724$\pm$0.030 & 0.725$\pm$0.024 \\
        4 & \checkmark & \checkmark & 35 & 143.1M &  0.722$\pm$0.041 & 0.724$\pm$0.033 \\
        5 & \checkmark & \checkmark & 21 & 169.5M &  0.716$\pm$0.043 & 0.719$\pm$0.039 \\
        6 & \checkmark & \checkmark & 7  & 193.8M &  0.718$\pm$0.039 & 0.720$\pm$0.031 \\
        
        \hline
    \end{tabular}
    \label{tab2}
\end{table*}

As shown in Table \ref{tab2}, we compare 6 types of instance generation methods. Compared to the results based on subnet pairs (21 instances), the results based on 7 subnets (7 instances) remarkably degrade, which strongly confirms the importance of using inter-subnet interactions in our model for disease diagnosis. Among the rest of the experiments, the one with subnet pairs (21 instances) reaches the best result. This could be reasonable because the instances covering more subnets may add difficulty to effectively mining the features. This is similar to the issue in whole-network-based feature extraction.

\begin{figure}[ht]
\includegraphics[width=\linewidth]{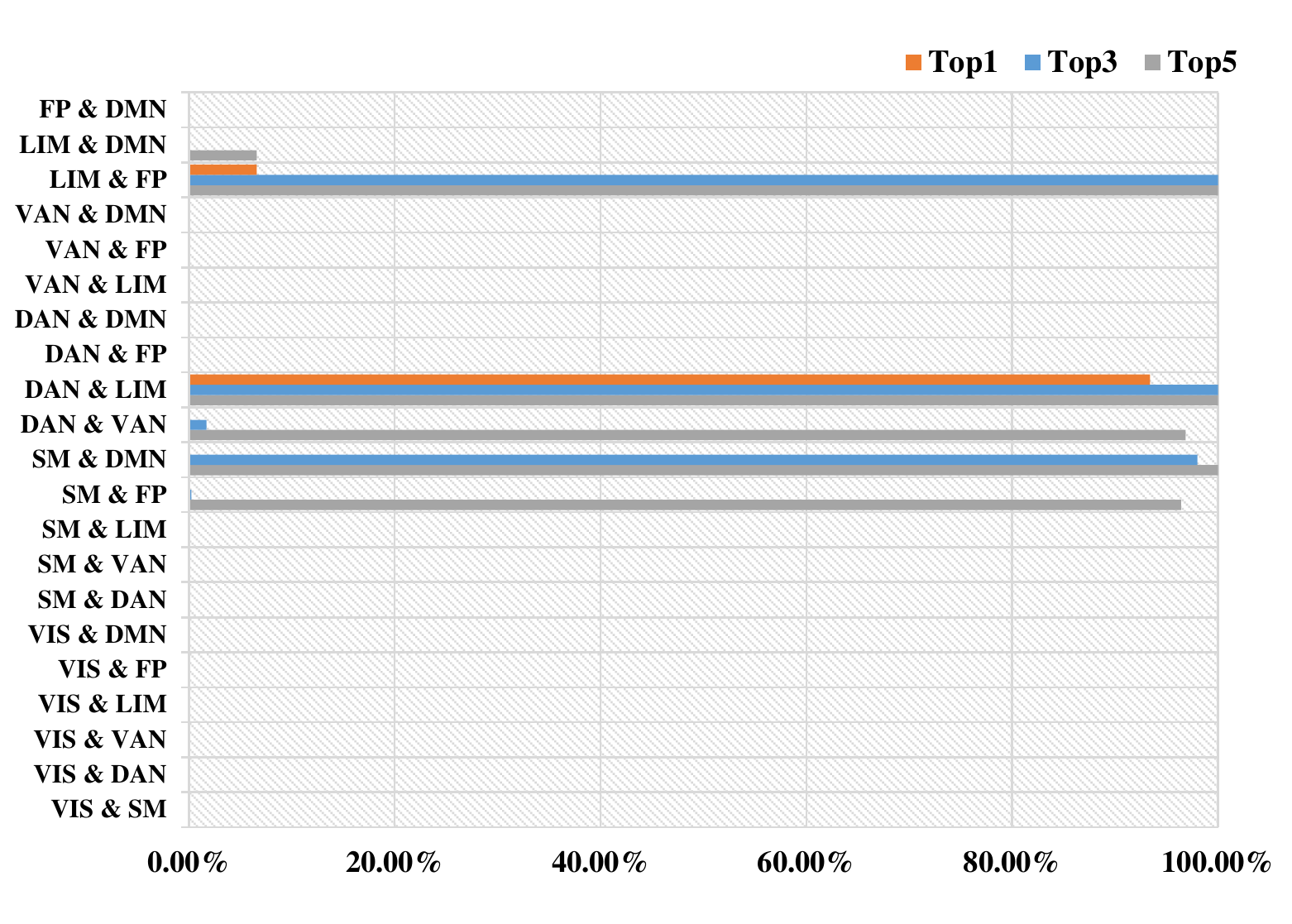}
\caption{The Top-K occurrence of subnet pairs. Details of subnet definition can be founded in \cite{schaefer2018local}: visual network (VIS), somatomotor network (SM), dorsal attention network (DAN), ventral attention network (VAN), limbic system (LIM), frontoparietal network (FP) and default mode network (DMN).} 
\label{fig2}
\end{figure}

\subsubsection{Feature Interpretability}

Based on the attention weights from our SLMIL strategy, we can identify the most predictive instance learned by our model. We test the best model (2-subnet pairs, with 21 instances) on all the samples, rank the attention weights, and count the occurrence of top-K for each instance (Fig.\ref{fig2}). We observe that the subnet between DAN and LIM is consistently identified (in 93.4\% cross-validation cases) as Top-1 predictive instance. This observation is in line with previous neurological studies \cite{zhang2015functional}\cite{bozoki2012disruption}. In addition, the LIM\&FP and SM\&DMN are consistently appearing within Top-3 predictive instances (nearly 100\% occurrence). A few other instances are identified. Note that the sparsity of predictive instance indicates that the diagnostic information is concentrated within a few subnet pairs.

\section{Conclusion}
\label{sec:conclusion}

In this work, we have proposed a novel graph learning framework for brain disease diagnosis, namely SLMIL-GCTrans. Through the experiments, our framework can achieve 72.9\% AUC and 73.1\% accuracy in differentiating EMCI from normal controls, which outperforms other compared methods. It is found that the pair of limbic system \& dorsal attention network is the most informative subnet pair in the diagnosis of EMCI. We expect that our method could also be generalized to other brain diseases for both diagnosis and the discovery of disease-related brain regions and connections.


\clearpage

\section{Compliance with Ethical Standards}
\label{sec:Compliance with Ethical Standards}
This research study is conducted retrospectively using human subject data made available in open access. Ethical approval is not required as confirmed by the license attached with the open access data.

\section{Acknowledgments}
\label{sec:Acknowledgments}
This work is supported in part by National Natural Science Foundation of China (grant number 62131015), and Science and Technology Commission of Shanghai Municipality
(STCSM) (grant number 21010502600).

\bibliographystyle{IEEEbib}
\bibliography{refs}

\end{document}